# Developing models to fit capacity-rate data in battery systems

Jonathan Coleman and Ruiyuan Tian

*School of Physics, CRANN and AMBER Research Centers, Trinity College Dublin, Dublin 2, Ireland*

*colemaj@tcd.ie (Jonathan N. Coleman); Tel: +353 (0) 1 8963859.

ABSTRACT: Analyzing rate-performance data in battery electrodes is greatly facilitated by access to simple analytical models. Here we describe a number of simple equations for fitting capacity-rate data. These equations output fit parameters, such as the characteristic time, which quantify rate-performance. This characteristic time can be linked to mechanistic effects, such as diffusion and electrical limitations, allowing quantitative analysis.

INTRODUCTION

Although research into lithium ion batteries has made rapid progress, high-rate-performance still needs to be significantly improved for a range of applications[1] such as rapid charging or high power delivery.[2] One factor hampering progress in this area is a lack of awareness among researchers that simple models exist to facilitate analysis of rate-performance data. Such models can be used to fit capacity-rate data, outputting fit parameters which enable the quantitative assessment of rate-performance and allow data analysis in terms of rate-limiting mechanisms.

In battery research the charge/discharge rate is denoted in various ways, most commonly via the specific charge/discharge current ($I/M$) or the C-rate, which is defined via C-rate $= C_R = (I/M)/(Q/M)_{Theory}$, where $(Q/M)_{Theory}$ is the theoretical capacity. More recently,[3] we have argued that the charge/discharge rate, R, defined as $R = (I/M)/(Q/M)$, where $Q/M$ represents the measured specific capacity, is an appropriate expression of rate where quantitative analysis is required.

The problem with rate-performance in batteries is based on the fact that, above some threshold charge/discharge rate (i.e. $(I/M)_T$, $R_T$ or $C_{R,T}$), the maximum achievable capacity begins to fall off with charge/discharge rate. Essentially, this limits the amount of energy a battery can deliver at high power, or store when charged rapidly. This is a significant problem and has led to a number of strategies targeting the electrodes,[4-7] electrolyte[8] and separator[9] with the aim of increasing $(I/M)_T$ and slowing the rate of capacity decay above $(I/M)_T$.

It is known that rate-performance can be improved by decreasing active particle size,[10-12] and electrode thickness,[13-16] or by increasing solid-state diffusivity,[10] conductor content[6,15,17] or electrode porosity[15,18] as well as by optimizing electrolyte concentration[13,15] and viscosity.[15] This implies that the factors limiting rate-performance are: solid-state diffusion of ions in the lithium-storing materials, electronic transport in electrodes, ion motion in both bulk electrolyte and electrolyte-filled pores and the timescale associated with electrochemical reactions.[11,19-21] Speeding up any of these processes should improve rate-performance.

However, for experimentalists, it is difficult to link the observed rate-performance to these factors quantitatively. The simplest and most common rate-performance data are capacity vs. rate curves (e.g. Q/M vs. C-rate). Although it is possible to use simple models to fit data and obtain information about the factors above, this is rarely done. Here we review some of the models available to analyze data, focusing on recently published, simple, analytical models. We note that in some cases, we have modified the equations from their reported form to make them consistent, in terms of form and notation, with the other equations we describe.

SIMULATIONS

A number of electrochemical models exist which fully describe the chemical processes occurring within the battery.[19,22-24] Most common is the Doyle-Fuller-Newman (DFN) approach which uses concentrated solution theory to model the charge/discharge processes in Li-ion cells.[25,26] These models involve the numerical solution of a six coupled differential equations which, once solved, yield a range of data including capacity as a function of current. Such models are comprehensive and match well to experimental data[13] but are relatively complex to use, making them inaccessible to the majority of experimentalists.

PHYSICS-BASED FITTABLE MODELS

Experimentalists require simple equations which can be used to fit capacity versus rate data, yielding fit parameters which can be linked to rate-limiting mechanisms. Here we present a number of such models, modifying them slightly to keep notation consistent. We note that areal (Q/A), volumetric (Q/V) and specific (Q/M) capacities are interchangeable via $Q/V = \rho_E \times Q/M$ and $Q/A = L_E \times Q/V$ where $\rho_E$ and $L_E$ are the electrode density and thickness. In addition, for simplicity, we normalize specific and volumetric capacities to the total mass and volume of the electrodes.

The earliest fittable equations were developed by dramatically simplifying the DFN model, resulting in three analytical models which describe capacity as a function of rate for three different rate-limiting processes: Ohmic limitations and diffusion in electrolyte or active particles.[21] However, in experimental systems, the rate-limiting process may not be known, raising questions as to which equation to use. As a result, these equations are not widely used for fitting purposes.

A considerably simpler approach was taken by Johns et al.[27] who proposed the capacity to be limited at high-rate by the lithium concentration in the electrolyte falling to zero within the electrode, leaving a "dead zone" where lithium cannot be stored. This leads to equations for Q/M as a function of either C-rate ($C_R$) or R:

$$\frac{Q}{M} = \frac{Q_M D_E C_0 F}{(1-t_+^0) Q_V^0 L_E^2 C_R} \qquad \text{or} \qquad \frac{Q}{M} = Q_M \left[ \frac{D_E C_0 F}{(1-t_+^0) Q_V L_E^2 R} \right]^{1/2} \qquad (1)$$

where $Q_M$ is the low-rate specific capacity, $Q_V$ and $Q_V^0$ are the low-rate experimental and theoretical volumetric capacities, $D_E$ is the diffusion coefficient of Li ions in the electrolyte within the pores of the electrode, $C_0$ is a measure of the concentration of Li ions, F is Faradays constant, $t_+^0$ is the Li ion transport number and $L_E$ is the electrode thickness. We note that the right-hand equation has the $R^{-1/2}$ dependence expected for a diffusion-limited process.[3]

Similarly, Gallagher et al.[22] used concentrated solution theory to infer the penetration depth of the electrolyte into the porous electrode. Applying this information results in expressions for specific capacity versus C-rate or R similar to equation 1:

$$\frac{Q}{M} = \frac{Q_M \gamma D_E C_0 F}{(1-t_+^0) Q_V^0 L_E^2 C_R} \qquad \text{or} \qquad \frac{Q}{M} = Q_M \left[ \frac{\gamma D_E C_0 F}{(1-t_+^0) Q_V L_E^2 R} \right]^{1/2} \qquad (2)$$

where $\gamma$ is the ratio of penetration length to electrode thickness. We note that equations 1 and 2 both apply only to the high-rate region where capacity falls with rate.

To address the latter point, Cornut et al.[28] proposed an equation which describes specific capacity as a function of C-rate for electrodes limited by solid state diffusion:

$$\frac{Q}{M} = \frac{Q_M}{1+(0.23 L_{AM}^2 / D_{AM}) \times C_R} \tag{3}$$

where $L_{AM}$ is the solid-state diffusion length associated with the active particles (related to particle size) and $D_{AM}$ is the diffusion coefficient of Li ions within the particles. This equation describes a capacity that is rate-independent at low-rate but decays inversely with C-rate at high-rate, and so is consistent with experimental data. However, it is limited to describing only one rate-limiting mechanism.

SEMIEMPIRICAL MODELS

One problem with the equations above is that they all give specific dependences on $C_R$, which may or may not describe real data. To address this, a number of authors have proposed semi-empirical equations to describe any capacity versus rate data. Of these, we reproduce three which have been proposed to express Q/M as a function of C-rate ($C_R$):

$$Q/M = Q_M \left(1 - \exp\left[-(R_C \tau)^{-n}\right]\right) \tag{4, Heubner et al)[29]}$$

$$Q/M = Q_M \exp\left[-(R_C \tau)^n\right] \tag{5, Wong et al)[30]}$$

$$Q/M = Q_M \left[1 - 2(\tau R_C)^n\right] \tag{6, Tian et al)[31]}$$

As well as two equations expressed in terms of charge/discharge rate, R:

$$\frac{Q}{M} = \frac{Q_M}{1 + 2(R\tau)^n} \tag{7, Tian et al.)[31]}$$

$$Q/M = Q_M \left[1 - (R\tau)^n \left(1 - e^{-(R\tau)^{-n}}\right)\right] \tag{8, Tian et al)[3]}$$

All of these equations fit data for Q/M as a function of $C_R$ or R very well, outputting fit parameters describing the low-rate capacity ($Q_M$), the characteristic charge/discharge time ($\tau$) and a constant (n). Probably most important parameter is the characteristic time. In all cases $\tau$ is a measure[3] of the rate where the capacity begins to fall off such that $1/\tau \propto (I/M)_T \propto R_T \propto C_{R,T}$ with the details depending on the specific capacity-rate equation. For example, if we define $R_T$ as the rate where the capacity falls below $0.9 \times Q_M$, equation 8 yields: $R_T = (1/10)^{1/n} / \tau$.[32]

The meaning of n can be illustrated by approximating the capacity-rate equations in the high-rate limit. This is most easily seen for equations 7 and 8, giving the same result in both cases: [3,31]

$$\left(\frac{Q}{M}\right)_{high-R} \approx \frac{Q_M}{2(R\tau)^n} \qquad (9)$$

meaning n is a measure of how fast the capacity decays with rate (c.f. equations 1 and 2).

Equations 4 to 8 have been plotted in figure 1 A-C (taking $\tau$=1 h, n=1) with all showing the familiar plateau at low rate followed by a decay at high-rates. Examples of fitting using these equations are given in figure 1D. We note that fitting data versus R rather than C-rate is probably advisable, as using the theoretical capacity to calculate C-rate (rather than the experimental capacity which is used to obtain R), can cause errors in $\tau$ when $Q_M<(Q/M)_{Theory}$ (see figure 1D). However, choice of fitting equation is largely a matter of taste. We recommend using equation 8, simply because databases of fit parameters found using this equation have been published.[3,33]

In their work, Heubner et al.[29] used eq 4 to fit a number of data sets, showing that the characteristic time, $\tau$, was similar to the timescale associated with ion diffusion in the electrolyte within the pores of the electrode:

$$\tau = \frac{L_E^2}{D_E} \qquad (10)$$

In a separate paper,[34] the same authors showed that a number of data sets can be explained by assuming that, similar to Johns et al.,[27] above a certain current, the Li ion concentration falls to zero within the electrode. This leads to a diffusion-limited (DL) current density, above which the capacity begins to decay. This current density can be converted to the diffusion-limited C-rate by dividing by $(Q/A)_{Theory}$. The authors showed that this DL C-rate matches well to the C-rate above which the capacity falls off, and so can be associated with $C_{R,T}$. Similarly, dividing the DL current density by the experimental Q/A (near the plateau, $Q/A \approx Q_V L_E$) gives $R_T$ in this scenario:

$$R_T \approx \frac{2D_E C_0 F}{Q_V L_E^2} \qquad (11a)$$

Equation 11a can be used in conjunction with equation 8 by applying $R_T = (1/10)^{1/n}/\tau$. Under these circumstances, this then yields a diffusion-limited characteristic time:

$$\tau = \frac{(1/10)^{1/n} Q_V L_E^2}{2 D_E C_0 F} \tag{11b}$$

Equation 11b shows some similarities to equation 10 but interestingly predicts the rate-performance to depend explicitly on the capacity. Comparison of equation 1 and equation 9 (taking n=1/2) shows that the work of Johns et al. is equivalent to a characteristic time of $\tau = Q_V L_E^2 (1-t_+^0)/(4 D_E C_0 F)$, which is very similar to equation 11b, as might be expected given the premises of these papers are equivalent. A very similar result can be found for equation 2.

Wong et al.[30] used equation 5 to fit a large number of Q/M v $C_R$ data sets from the literature. They found correlations between $\tau$ and various electrode parameters such as particle size and the activation energy associated with solid state diffusion (with the latter relationship implying an inverse scaling between $\tau$ and solid-state diffusion coefficient, consistent with equation 3 and equation 12 below).

Equations 6 and 7 were both derived[31] using methods[35] designed to convert chronoamperometry data to capacity-rate data. These equations have the advantage that they have some physical basis, are expressed in a simple form and fit experimental data well. In addition, both of these equations were designed to output values of $Q_M$, $\tau$ and n which are consistent with published values obtained from equation 8.

Tian et al.[3] used equation 8 to fit ~200 capacity-rate data sets extracted from the literature (figure 2A). Interestingly, they performed detailed analysis on the fit parameters, focusing on n and $\tau$. By plotting a histogram of the obtained n-values, they found this value to fall roughly between 0.5 and 1, the values expected for diffusion-limited and electrically-limited behaviour (figure 2B).[3,36] They also found a rough scaling of $\tau$ with $L_E^2$ over the whole data set (figure 2C). This result is particularly important as it shows the importance of electrode thickness and illustrates that rate-performance cannot be reported without reference to electrode thickness. This scaling implies that the quantity $L_E^2/\tau$ can be used as a figure of merit for rate-performance. This parameter varied between ~$10^{-14}$ m$^2$/s and ~$10^{-9}$ m$^2$/s for very fast electrodes, yielding a scale to which rate-performance can be compared (figure 2D).

MULTIPLE RATE-LIMITING MECHANISMS

Tian et al.[3] proposed that all possible rate-limiting mechanisms could accounted for by writing $\tau$ as a combination of the RC charge/discharge time of the electrode, various diffusion times and the time associated with the electrochemical reaction, $t_c$:

$$\tau = L_E^2 \left[ \frac{C_{V,eff}}{2\sigma_{OOP}} + \frac{C_{V,eff}}{2\sigma_{BL} P_E^{3/2}} + \frac{1}{D_{BL} P_E^{3/2}} \right] + L_E \left[ \frac{L_S C_{V,eff}}{\sigma_{BL} P_S^{3/2}} \right] + \left[ \frac{L_S^2}{D_{BL} P_S^{3/2}} + \frac{L_{AM}^2}{D_{AM}} + t_c \right] \qquad (12)$$

Here $C_{V,eff}$ is the effective volumetric *capacitance* of the electrode, $\sigma_{OOP}$ is the out-of-plane electrical conductivity of the electrode, $P_E$ and $P_S$ are the porosities of the electrode and separator respectively while $L_S$ is the separator thickness. Here $\sigma_{BL}$ is the overall (anion and cation) conductivity of the bulk electrolyte (S/m) while $D_{BL}$ is the cation diffusion coefficient in the bulk electrolyte. We note that this equation includes the effect of equation 10 (3rd term in equation 12) and the bracketed part of equation 3 (6th term). In addition, it can be shown that if the Nearnst-Einstein equation is applied to equation 11b, it becomes very similar to the 2nd term in equation 12 (once $C_{V,eff}$ is replaced using the empirical relation below).

Equation 12 predicts $\tau$ to scale quadratically with $L_E$. Tian et al showed this to be true for a range of materials[3] as illustrated[31] in figure 3A. The authors tested the predications of this equation against experimental data in a number of ways, finding quantitative matches in all cases.[3] For example, as shown in figure 3B, the predicted dependence of $\tau$ on separator thickness is found experimentally.[3] In addition, by analysing electrode thickness-dependent data, Tian et al. found an empirical relationship between $C_{V,eff}$ and the low-rate volumetric capacity of the electrode, $Q_V$, such that: $C_{V,eff} / Q_V = 28$ F/mAh (figure 3C), a relationship that facilitates quantitative analysis of data.[3]

Tian at al.[36] followed this work by characterising the rate-performance of electrodes as a function of electrode conductivity, which they controlled by varying the conductive additive content. Combining equation 12 with the empirical equation for $C_{V,eff}$ given above predicts that:

$$\frac{\tau}{L_E^2} = 14 \frac{Q_V}{\sigma_{OOP}} + \frac{\beta}{L_E^2} \qquad (13)$$

Where β is just the sum of terms 2-7 in equation 12. Tian et al. showed that this equation matches experimental data almost exactly. This strongly supports the veracity of both equation 12 and the empirical relation obtained from figure 3C.

CONCLUSIONS

This review demonstrates that a number of simple models exist that allow the fitting of capacity-rate data in battery systems. Such models range from single-mechanism equations which only fit high-rate behaviour to semi-empirical models which fit the entire rate range, outputting parameters which allow the quantitative assessment of rate-behaviour. Of such parameters, the characteristic time is the most important as it can be linked to rate-limiting mechanisms, allowing quantitative analysis.

**Conflicts of interest statement**: Nothing declared.

**Acknowledgments:** The authors acknowledge the SFI-funded AMBER research centre (SFI/12/RC/2278) and Nokia for support. JNC thanks Science Foundation Ireland (SFI, 11/PI/1087) and the Graphene Flagship (grant agreement n°785219) for funding.

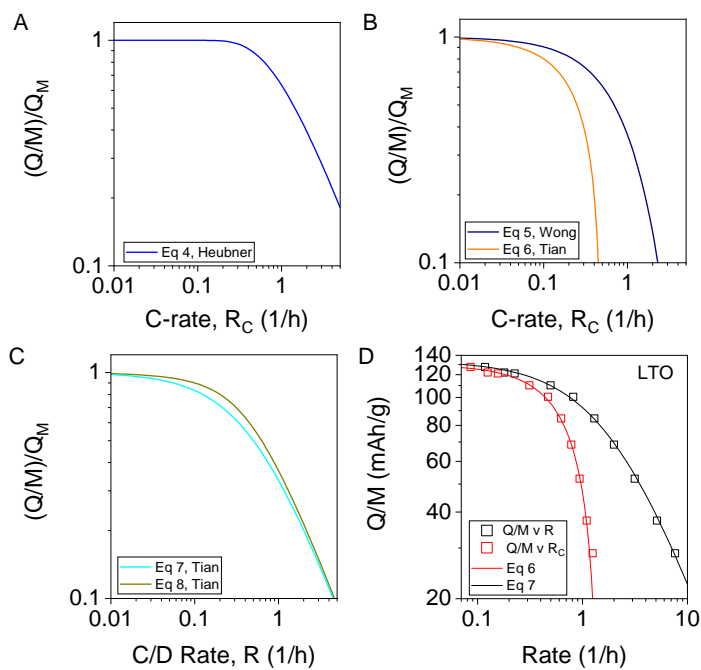

Figure 1: Plots of equation 4 (A), equations 5 and 6 (B) and equations 7 and 8 (C) versus either C-rate ($R_C$) or charge/discharge rate (R). In each case, $\tau$=1 h and n=1. We note that the curve in A is more similar in shape to those in C than those in B. As such we have proposed that equation 4 should be written in terms of R rather than C-rate.[31] D) Examples of using such equations to fit capacity versus rate data.[31] Q/M is plotted versus both $R_C$ and R for an NMC electrode. These equations are fit to equations 6 and 8 respectively. The outputted fit parameters are: Q/M vs. $R_C$ (red) – $Q_M$=130 mAh/g, $\tau$=0.40 h, n=1.25; Q/M vs. R (black) – $Q_M$=134 mAh/g, $\tau$=0.24 h, n=1.04. The $R_C$ values were calculated taking $(Q/M)_{Theory}$=175 mAh/g. Because $Q_M$<$(Q/M)_{theory}$, the C-rate fits do not give accurate values of $\tau$.

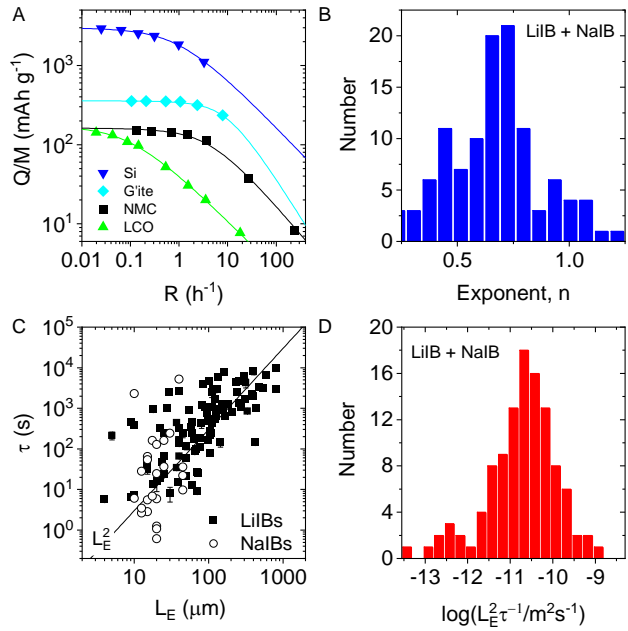

Figure 2: Fitting capacity-rate data and analysing fit parameters. A) Examples of Q/M vs. R curves for four electrode materials fitted using equation 8. B) Histogram of n-values found from fitting published Q/M vs. R curves. C) Plot of characteristic time, τ, extracted from fitting published data plotted versus electrode thickness, $L_E$. The line indicates $\tau \propto L_E^2$. B) Histogram of values $L_E^2/\tau$ found from fitting published Q/M vs. R curves. Panels B-D include data for both lithium and sodium ion batteries. This data was adapted from ref[3].

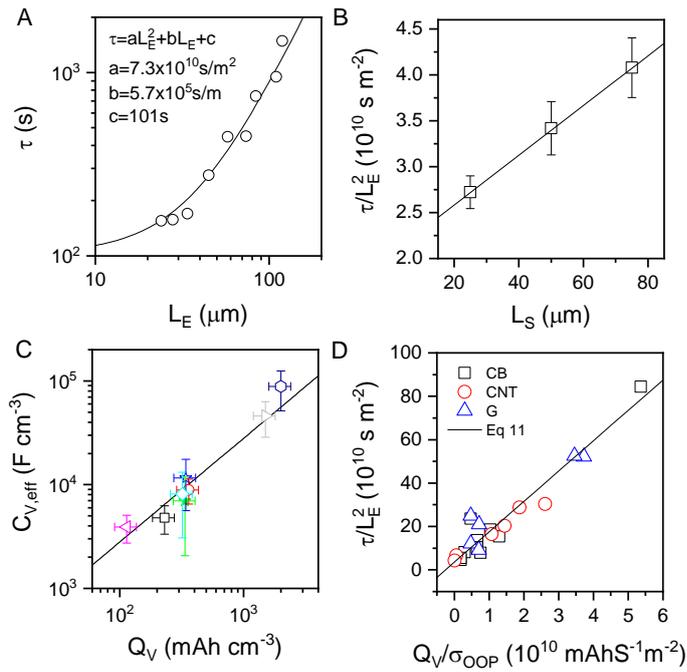

Figure 3: Analysing τ and related data. A) Characteristic time, τ, found by fitting capacity-rate data for NCA electrodes,[31] plotted versus electrode thickness, $L_E$. The line is a fit to equation 12. B) Values of $\tau/L_E^2$ for NMC electrodes[3] plotted versus separator thickness. The solid-line is predicted by equation 12. C) Empirical relationship between effective volumetric capacity and volumetric capacitance. The line is consistent with $C_{V,eff}/Q_V = 28 F/mAh$ [3]. D) Values $\tau/L_E^2$ for NMC electrodes[36] filled with various mass fractions of carbon black, nanotubes or graphene (and so with different out-of-plane conductivities, $\sigma_{OOP}$) plotted versus $Q_V/\sigma_{OOP}$. The solid line is a plot of equation 13, taking reasonable values for each parameter (using equation 12 to find β).[36]

References

1. Marom R, Amalraj SF, Leifer N, Jacob D, Aurbach D: **A review of advanced and practical lithium battery materials**. *Journal of Materials Chemistry* 2011, **21**:9938-9954, 10.1039/c0jm04225k.
2. Eftekhari A: **Lithium-Ion Batteries with High Rate Capabilities**. *Acs Sustainable Chemistry & Engineering* 2017, **5**:2799-2816, 10.1021/acssuschemeng.7b00046.
3. Tian R, Park SH, King PJ, Cunningham G, Coelho J, Nicolosi V, Coleman JN: **Quantifying the factors limiting rate performance in battery electrodes**. *Nat Commun* 2019, **10**:1933, 10.1038/s41467-019-09792-9.
4. Ge H, Chen L, Yuan W, Zhang Y, Fan Q, Osgood H, Matera D, Song X-M, Wu G: **Unique mesoporous spinel Li 4 Ti 5 O 12 nanosheets as anode materials for lithium-ion batteries**. *Journal of Power Sources* 2015, **297**:436-441, 10.1016/j.jpowsour.2015.08.038.

5. Kang B, Ceder G: **Battery materials for ultrafast charging and discharging**. *Nature* 2009, **458**:190-193, 10.1038/nature07853.
6. Zhang CF, Park SH, Ronan O, Harvey A, Seral-Ascaso A, Lin ZF, McEvoy N, Boland CS, Berner NC, Duesberg GS, et al.: **Enabling Flexible Heterostructures for Li-Ion Battery Anodes Based on Nanotube and Liquid-Phase Exfoliated 2D Gallium Chalcogenide Nanosheet Colloidal Solutions**. *Small* 2017, **13**,
7. Liu YP, He XY, Hanlon D, Harvey A, Khan U, Li YG, Coleman JN: **Electrical, Mechanical, and Capacity Percolation Leads to High-Performance MoS2/Nanotube Composite Lithium Ion Battery Electrodes**. *Acs Nano* 2016, **10**:5980-5990, 10.1021/acsnano.6b01505.
8. R. LE, M. TE, L. GK, Jing L, Xiaowei M, Y. BL, R. DJ: **A Study of the Physical Properties of Li-Ion Battery Electrolytes Containing Esters**. *Journal of the Electrochemical Society* 2018, **165**:A21-A30,
9. Zhang B, Wang QF, Zhang JJ, Ding GL, Xu GJ, Liu ZH, Cui GL: **A superior thermostable and nonflammable composite membrane towards high power battery separator**. *Nano Energy* 2014, **10**:277-287, 10.1016/j.nanoen.2014.10.001.
10. Du WB, Gupta A, Zhang XC, Sastry AM, Shyy W: **Effect of cycling rate, particle size and transport properties on lithium-ion cathode performance**. *International Journal of Heat and Mass Transfer* 2010, **53**:3552-3561, 10.1016/j.ijheatmasstransfer.2010.04.017.
11. Ye JC, Baumgaertel AC, Wang YM, Biener J, Biener MM: **Structural Optimization of 3D Porous Electrodes for High-Rate Performance Lithium Ion Batteries**. *Acs Nano* 2015, **9**:2194-2202, 10.1021/nn505490u.
12. Xue L, Li XP, Liao YH, Xing LD, Xu MQ, Li WS: **Effect of particle size on rate capability and cyclic stability of LiNi0.5Mn1.5O4 cathode for high-voltage lithium ion battery**. *Journal of Solid State Electrochemistry* 2015, **19**:569-576, 10.1007/s10008-014-2635-4.
13. Doyle M, Newman J, Gozdz AS, Schmutz CN, Tarascon JM: **Comparison of modeling predictions with experimental data from plastic lithium ion cells**. *Journal of the Electrochemical Society* 1996, **143**:1890-1903, 10.1149/1.1836921.
14. Zhao R, Liu J, Gu JJ: **The effects of electrode thickness on the electrochemical and thermal characteristics of lithium ion battery**. *Applied Energy* 2015, **139**:220-229, 10.1016/j.apenergy.2014.11.051.
15. Yu DYW, Donoue K, Inoue T, Fujimoto M, Fujitani S: **Effect of electrode parameters on LiFePO4 cathodes**. *Journal of the Electrochemical Society* 2006, **153**:A835-A839,
16. Zheng HH, Li J, Song XY, Liu G, Battaglia VS: **A comprehensive understanding of electrode thickness effects on the electrochemical performances of Li-ion battery cathodes**. *Electrochimica Acta* 2012, **71**:258-265,
17. Zhang B, Yu Y, Liu YS, Huang ZD, He YB, Kim JK: **Percolation threshold of graphene nanosheets as conductive additives in Li4Ti5O12 anodes of Li-ion batteries**. *Nanoscale* 2013, **5**:2100-2106,
18. Bauer W, Notzel D, Wenzel V, Nirschl H: **Influence of dry mixing and distribution of conductive additives in cathodes for lithium ion batteries**. *Journal of Power Sources* 2015, **288**:359-367,
19. Jiang FM, Peng P: **Elucidating the Performance Limitations of Lithium-ion Batteries due to Species and Charge Transport through Five Characteristic Parameters**. *Scientific Reports* 2016, **6**, 10.1038/srep32639.
20. Zhang HG, Yu XD, Braun PV: **Three-dimensional bicontinuous ultrafast-charge and -discharge bulk battery electrodes**. *Nature Nanotechnology* 2011, **6**:277-281, 10.1038/nnano.2011.38.
21. Doyle M, Newman J: **Analysis of capacity-rate data for lithium batteries using simplified models of the discharge process**. *Journal of Applied Electrochemistry* 1997, **27**:846-856, 10.1023/a:1018481030499.
22. Gallagher KG, Trask SE, Bauer C, Woehrle T, Lux SF, Tschech M, Lamp P, Polzin BJ, Ha S, Long B, et al.: **Optimizing Areal Capacities through Understanding the Limitations of Lithium-Ion Electrodes**. *Journal of the Electrochemical Society* 2016, **163**:A138-A149,

23. Danner T, Singh M, Hein S, Kaiser J, Hahn H, Latz A: **Thick electrodes for Li-ion batteries: A model based analysis**. *Journal of Power Sources* 2016, **334**:191-201, 10.1016/j.jpowsour.2016.09.143.
24. Schmidt AP, Bitzer M, Imre AW, Guzzella L: **Experiment-driven electrochemical modeling and systematic parameterization for a lithium-ion battery cell**. *Journal of Power Sources* 2010, **195**:5071-5080, 10.1016/j.jpowsour.2010.02.029.
25. Doyle M, Fuller TF, Newman J: **Modeling of Galvanostatic Charge and Discharge of the Lithium Polymer Insertion Cell**. *Journal of the Electrochemical Society* 1993, **140**:1526-1533, 10.1149/1.2221597.
26. Fuller TF, Doyle M, Newman J: **Simulation and Optimization of the Dual Lithium Ion Insertion Cell**. *Journal of the Electrochemical Society* 1994, **141**:1-10, 10.1149/1.2054684.
27. Johns PA, Roberts MR, Wakizaka Y, Sanders JH, Owen JR: **How the electrolyte limits fast discharge in nanostructured batteries and supercapacitors**. *Electrochemistry Communications* 2009, **11**:2089-2092, 10.1016/j.elecom.2009.09.001.
28. Cornut R, Lepage D, Schougaard SB: **Interpreting Lithium Batteries Discharge Curves for Easy Identification of the Origin of Performance Limitations**. *Electrochimica Acta* 2015, **162**:271-274, 10.1016/j.electacta.2014.11.035.
29. Heubner C, Seeba J, Liebmann T, Nickol A, Borner S, Fritsch M, Nikolowski K, Wolter M, Schneider M, Michaelis A: **Semi-empirical master curve concept describing the rate capability of lithium insertion electrodes**. *Journal of Power Sources* 2018, **380**:83-91, 10.1016/j.jpowsour.2018.01.077.
30. Wong LL, Chen HM, Adams S: **Design of fast ion conducting cathode materials for grid-scale sodium-ion batteries**. *Physical Chemistry Chemical Physics* 2017, **19**:7506-7523, 10.1039/c7cp00037e.
31. Tian R, King PJ, Coelho J, Park S-H, Nicolosi V, O'Dwyer C, Coleman JN: **Using chronoamperometry to rapidly measure and quantitatively analyse rate performance in battery electrodes**. *ArXiv 1911.12305* 2019,
32. Park SH, Tian RY, Coelho J, Nicolosi V, Coleman JN: **Quantifying the Trade-Off between Absolute Capacity and Rate Performance in Battery Electrodes**. *Advanced Energy Materials* 2019, **9**:10, 10.1002/aenm.201901359.
33. Tian R, Breshears M, Horvath DV, Coleman JN: **Do 2D material-based battery electrodes have inherently poor rate-performance?** *ArXiv: 1912.02482* 2019,
34. Heubner C, Schneider M, Michaelis A: **Diffusion-Limited C-Rate: A Fundamental Principle Quantifying the Intrinsic Limits of Li-Ion Batteries**. *Advanced Energy Materials*, 10.1002/aenm.201902523.
35. Heubner C, Lammel C, Nickol A, Liebmann T, Schneider M, Michaelis A: **Comparison of chronoamperometric response and rate-performance of porous insertion electrodes: Towards an accelerated rate capability test**. *Journal of Power Sources* 2018, **397**:11-15, 10.1016/j.jpowsour.2018.06.087.
36. Tian R, Alcala N, O'Neill SJ, Horvath D, Coelho J, Griffin A, Zhang Y, Nicolosi V, O'Dwyer C, Coleman JN: **Quantifying the effect of electrical conductivity on the rate-performance of nanocomposite battery electrodes** *ArXiv:1912.02485* 2019,

Papers of special interest (*) or outstanding interest (**) [to be inserted after acceptance when endnote is disabled]

3** (Tian, Nat Comm) Introduces a semiempirical equation for fitting capacity versus rate and applied it to 200 data dets from the literature. Analyses resultant n and τ data. Proposes an equation describing τ in terms of all rate-limiting mechanisms.

22* (Gallagher, JES) Proposes an equation for penetration depth, leading for an equation for capacity as a function of rate.

27* (Johns, EC) Uses penetration depth to propose equation for capacity as a function of rate.

28* (Cornut, EA) Proposes a capacity-rate equation for electrodes limited by solid state diffusion.

29** (Heubner, JPS) Proposes a semiempirical capacity-rate equation and demonstrates the characteristic time is limited by electrolyte diffusion in pores.

30** (Wong, PCCP) Proposes a semiempirical capacity-rate equation and demonstrates how the characteristic time scales with electrode properties.

31* (Tian, ArXiv) Derives equations for capacity as a function of both R and C-rate.

34* (Heubner, AEM) Introduces diffusion-limited C-rate as rate-limiting factor.

35** (Heubner, JPS) Shows chronoamperometry can be used to accelerate rate testing.